# It is Rotating Leaders Who Build the Swarm: Social Network Determinants of Growth for Healthcare Virtual Communities of Practice

Antonacci, G., Fronzetti Colladon, A., Stefanini, A. & Gloor, P.A.

# It is Rotating Leaders Who Build the Swarm: Social Network Determinants of Growth for Healthcare Virtual Communities of Practice

Antonacci, G., Fronzetti Colladon, A., Stefanini, A. & Gloor, P.

**Abstract**

**Purpose:** This 7-year longitudinal study identifies factors influencing the growth of healthcare Virtual Communities of Practices (VCoPs) using metrics from social-network and semantic analysis. Studying online communication along the three dimensions of social interactions (connectivity, interactivity and language use) we aim to provide VCoPs managers with valuable insights to improve the success of their communities.

**Design/Methodology:** Communications over a period of 7 years (April 2008 to April 2015), and between 14,000 members of 16 different healthcare VCoPs coexisting on the same web-platform, were analyzed. Multilevel regression models were used to reveal the main determinants of community growth over time. Independent variables were derived from social network and semantic analysis measures.

**Findings:** Results show that structural and content-based variables predict the growth of the community. Progressively more people will join a community if: its structure is more centralized, leaders are more dynamic (they rotate more), and the language used in the posts is less complex.

**Research limitations/implications:** The available dataset included one web platform and a limited number of control variables. In order to consolidate the findings of the present study, the experiment should be replicated on other healthcare VCoPs.

**Originality/value:** The study provides useful recommendations for setting up and nurturing the growth of professional communities, considering at the same time the structure of the interaction patterns among community members, the dynamic evolution of these interactions and the use of language. New analytical tools are presented, together with the use of innovative interaction metrics which can significantly influence community growth, such as rotating leadership.

# 1. Introduction

*Virtual Communities of Practices* (VCoPs) in healthcare are becoming increasingly popular as technology enables and facilitates knowledge exchange among health professionals from all over the world, including those areas which are geographically or professionally isolated (Dieleman and Duncan, 2013). In addition to websites which more generally target professionals of any industry (e.g. LinkedIn), an increasing number of social networking platforms which are dedicated to healthcare professionals exist. This offers healthcare professionals the possibility of collaborating with their peers, to manage information and to develop and spread knowledge (Jennings Mabery, Gibbs-Scharf, and Bara, 2013; Mairs et al., 2013).

The rapid growth in dimension and popularity of VCoPs is not surprising (Dieleman and Duncan, 2013; Li et al., 2009; Ho et al., 2010) as social networking theories point out how knowledge sharing and learning in the digital era have mainly become a social activity (Pettenati and Cigognini, 2007). Joining VCoPs has therefore become a great opportunity for highly knowledge-intensive and specialized healthcare professionals.

Thanks to VCoPs, healthcare professionals can access a common platform where they can learn, discuss or share resources with colleagues, thus building a professional support network and accelerating the process of translation of evidence-based research into daily practice (Rolls et al., 2008; Parboosingh, 2002; Winkelman and Choo, 2003, Ho et al., 2010). VCoPs in healthcare can be used to discuss problems, best practice, work on research projects by sharing resources or generating new ideas, thus obtaining benefits in terms of increased quality of care, creation of new knowledge, as well as increased productivity (Demiris, 2006; Endsley, Kirkegaard, and Linares, 2005; Robinson and Cottrell, 2005; Sandars and Heller, 2006). Moreover, the opportunity to share knowledge overcoming geographical barriers proved to support both personal professional development, and the overall performance of the process of care delivery (Dieleman and Duncan, 2013).

Factors favoring knowledge-sharing in VCoPs have long been in the focus of researchers (Aljuwaiber, 2016; Hara and Hew, 2007; Stewart and Abidi, 2012), who, in addition, have highlighted that the successful design and management strategies of online communities heavily depend also on the community's specific characteristics (Andrews, 2002; Li et al., 2009; Iriberri and Leroy, 2009). However, although the role played by VCoPs in healthcare as enabler and facilitator of knowledge exchange among health professionals, policy makers, and researchers, is widely recognized as pivotal (Mairs et. al. 2013), there are still only a few studies examining the characteristics of online platforms in this area (Ho et al., 2010; Hara and Hew, 2007; Stewart and Abidi, 2012). Moreover, even though previous studies recognized social dynamics as one of the most important determinants of members' participation in VCoPs (Amichai-Hamburger et al., 2016; Hara and Hew, 2007), little documented research has analyzed participation in VCoPs from the perspective of social interaction and group processes (Malinen, 2015).

The purpose of this study is twofold. On the one side, we aim to extend the theoretical body of research on VCoPs by contributing to a greater understanding of social processes between community members and their impact on the growth of healthcare VCoPs. On the other side, we aim to provide VCoPs managers with useful insights to build and sustain the flourishing of their communities, in particular through the use of online-platform analytics tools.

Most studies about social processes in VCoPs (e.g., Hara and Hew, 2007; Nistor et al., 2014; Stewart and Abidi, 2012) focused on the analysis of structural metrics using Social Network Analysis (SNA) (Wasserman and Faust, 1994; Malinen, 2015) and disregarded other relevant aspects of social interactions, such as longitudinal dynamics, sentiment or emotionality of the language used. To the best of our knowledge, there is no research investigating the joint effects of community structure, longitudinal interactions and language used, on the growth of VCoPs. To this aim, we conducted a 7-year longitudinal study on 16 healthcare VCoPs coexisting on a common web platform. We carried out our quantitative analysis along the three dimensions of social interaction: use of language, connectivity (structural analysis), and interactivity (longitudinal analysis of the interactions) (Gloor et al., 2017a). Although the present study has been set in the healthcare domain, the general results we present might be tested and eventually extended to VCoPs operating in other professional domains.

## 2. Virtual Communities of Practice

Advances in technology are deeply changing the way people communicate and exchange knowledge. The rise of the Internet started many forms of online sociality, including e-mail, instant messaging, blogging, social networking and other online services (Heer and Boyd, 2005). In the last decades, social interaction has progressively become the most popular activity among Internet users, and virtual communities (VCs) have emerged and rapidly increased in number and size (Zickuhr, 2010; Gross, 1999; Petersen, 1999). VCs can be described as "social aggregations that emerge from the Net when enough people carry on public discussions long enough, with sufficient human feeling" (Rheingold, 1993, p. 5). The main purposes of VCs are knowledge sharing and social interaction (Burnett, 2000; Ridings and Gefen, 2004). VCs are characterized by a persistent interaction between members (Smith, 1999; Ridings, Gefen, and Arinze, 2002), which makes them feel part of a greater social group, establishes a notion of membership, and contributes to the development of relationships among participants (Figallo, 1998; Sproull and Faraj, 1997). Compared to face-to-face communities, electronic means make participation independent from the physical location and allow a larger and a more heterogeneous composition of joiners (Sproull and Faraj, 1997), thus facilitating communication and enhancing diversity inside the learning network. Allen and colleagues (2016) proved the value of communication regardless of physical proximity: the quality of the exchanged knowledge and the communication behaviors has a much stronger impact than physical clustering. The characteristics provided by virtual interaction lead to the development of "weak tie" relationships (Granovetter, 1973), making VCs an ideal place to exchange knowledge and information also with people not known in person (Constant, Sproull, and Kiesler, 1996). Consistently, knowledge exchange is one of the main motivations leading people to participate in VCs (Wasko and Faraj, 2000).

There are various types of online communities (Lazar and Preece, 1998; Gain, 1997; Stewart, 2010), each built with a different purpose and providing specific benefits to its members. In particular, online communities include Communities of Practice (CoPs) for professionals, enterprise communities, social network websites, wikis, creative and question-answer sites (Malinen, 2015). The so-called virtual CoPs (VCoPs) are a kind of VC.

The Community of Practice (CoP) is still an evolving concept (Li et al., 2009). Several definitions of CoP have been given by different authors (Lave and Wenger, 1991; Brown and Duguid, 1991; Wenger, 1998; Wenger, McDermott, and Snyder, 2002). While definitions differ from each other because they focus on different perspectives from which CoPs can be analyzed (Cox, 2005), all of them agree on the key role played by social interaction to foster learning and knowledge sharing among individuals. From the perspective of our study, we refer to COPs as: "… groups of people who share a concern, a set of problems, or a passion about a topic, and who deepen their knowledge and expertise in this area by interacting on an ongoing basis" (Wenger, McDermott, and Snyder, 2002, p. 4). Wenger, McDermott, and Snyder (2002) outlined three main elements characterizing CoPs: *Domain*, *Community* and *Practice*. VCoPs' members participate in a process of collective learning through social interactions (Community) within a specific domain of interest representing the "common ground" (Domain) on which they develop, share and maintain knowledge (Practice) (Wenger, McDermott, and Snyder, 2002). These three elements should be all present and work well together in order to nurture the process of creation and dissemination of knowledge within a VCoP (Wenger, McDermott, and Snyder, 2002; Wenger and Snyder, 2000; Wenger, 2000).

*2.2 Improving participation in Communities of Practice*

While some communities exhibit extraordinary membership growth, there are many others, which languish or rapidly vanish. Recent studies show that by 2010 the majority of the Fortune 1000 companies experimented with some form of online communities and that such efforts failed to reach the desired outcome in more than half of the cases (Sarner, 2008). Market data show that 70% of online communities fail (Gartner, 2012). It cannot be assumed that people will automatically join and actively participate in a VC. Even when people have common interests or objectives, they may have little or no intention to share it on a social network (Andrews, 2002).

In response, researchers from various disciplines are searching for conditions, strategies and guidelines to face the challenge of building and sustaining successful VCs. As a result, a wide body of theoretical and empirical literature has been developed to derive or empirically test drivers and contextual factors which are likely to play an important role in the growth of online communities (Paroutis and Al Saleh, 2009; Iskoujina and Roberts, 2015; Barker, 2015; Leimeister, Ebner, and Kremar, 2005; Kraut et al., 2012; Hall and Graham, 2004; Kim, 2000; Dubé, Bourhis, and Jacob, 2005; Preece, 2000; Kling and Courtright, 2003). In particular, two major research streams emerged in this field: (i) research focusing on the analysis of the steps related to creating a successful platform with active participation (e.g. Andrews, 2002; Matei et al., 2015); and (ii) research aimed to identify and describe the important drivers of popularization of existing platforms (e.g. Gunawardena et al., 2009; Spagnoletti, Resca, and Lee, 2015).

Factors that motivate and hinder participation within online communities were investigated by many studies, which focus on different aspects (Koh et al., 2007) and point out that participation in online communities can either be passive or active (Malinen, 2015, Jones, Ravid, and Rafaeli, 2004; Nonnecke and Preece, 2001). While active members actively contribute to the community, passive members participate in the community taking advantage of the knowledge offered without contributing to the community activities. In our empirical research, we focus on social interactions among factors that influence active participation.

Participation represents an essential factor for the flourishing of VCoPs as it leads to the accumulation of experience and stimulates the social construction of knowledge and the development of expertise (Bishop, 2007; Paavola, Lipponen, and Hakkarainen, 2004; Nistor et al., 2014). VCoPs are characterized by self-selection and the voluntary nature of participation, as people voluntarily choose to contribute to joint activities and discussions, help other members and share information (Wasko, Faraj, and Teigland, 2004). Therefore, VCoPs exist as long as participation to the community is of value to their members (Gray, 2004) and the value of a VCoP largely depends on the ongoing participation of its members (Chen, 2007; Fang and Chiu, 2010; Cheung, Lee, and Lee, 2013).

Literature about online knowledge sharing and literature on VCoPs point out three macro-factors affecting the participation in online communities: (i) individual factors, (ii) technological factors and (iii) social factors (Hara and Hew, 2007; Malinen, 2015; Amichai-Hamburger, 2016; Paroutis and Al Saleh, 2009; Ardichvili, Page, and Wentling, 2003; Nistor et al., 2014). While *technological factors* involve technical and usability issues about online communities, *individual factors* refer to CoPs members' characteristics such as motivations, personalities, values, time available and other individual traits. *Social factors*, on the other hand, are related to the social-group processes and include aspects such as the role of individuals within the community (e.g. moderator roles) and the social interaction among members, posting news or questions and answering one another.

This study is focused on factors influencing the growth of healthcare VCoPs. The framework underlying our research is presented in Figure 1.

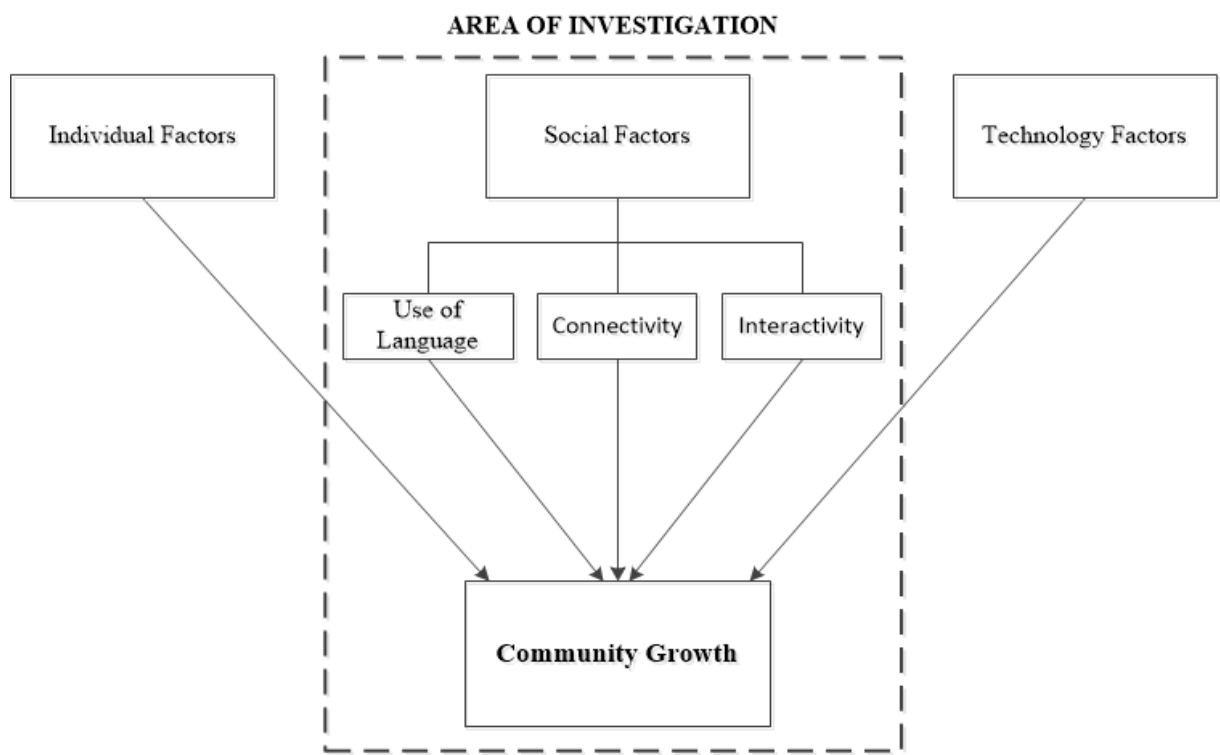

**Figure 1**. Theoretical Framework.

We focus on social factors for different reasons. Firstly, behaviors and communication patterns in the social environment, have been found to play a pivotal role in the development of CoPs and VCoPs (Amichai-Hamburger et al., 2016; Battistoni and Fronzetti Colladon, 2014; Jeon, Kim and Koh, 2011; Hara and Hew, 2007; Nistor et al., 2014). Secondly, even though technological factors are recognized as important, they are not the main determinants for abandoning or preventing participation (Maloney-Krichmar and Preece, 2005; Brandtzæg and Heim, 2008).

The impact of social processes on the success of online communities has been investigated by different perspectives and using either qualitative methods, quantitative methods or both (Malinen, 2015; Liao and Chou, 2012; Dennen, 2014). As our objective was to find success drivers for VCoPs, which could be easily assessed and monitored also by managers of online platforms with a high volume of members and information exchange, in this study we adopt a quantitative approach. In this context, SNA is recognized as a valuable investigating method (Malinen, 2015) and has been used in many previous studies to analyze the social interactions inside online communities (Zaphiris and Sarwar, 2006; Pfeil and Zaphiris, 2009; Stewart and Abidi, 2012; Nistor et al., 2014). However, these studies focused mainly on the volume of communication and on the density and patterns of the network (Malinen, 2015). We maintain that a purely structural approach to social relations is not sufficient to explore the social processes underlying online community dynamics (Yuan, 2013; Malinen, 2015) – a more comprehensive understanding of social interactions requires content/sentiment metrics to be studied alongside network centrality measures.

Accordingly, we propose to analyze social processes adopting the conceptualization of social interactions and correspondent metrics proposed by Gloor and colleagues (2006, 2016, 2017a, 2017b). Their approach is the result of research conducted over the last fourteen years at the MIT Center for Collective Intelligence: hundreds of communications archives (Twitter, Facebook, Wikipedia, e-mail and face-to-face network) were analyzed and key characteristics of high performing groups were identified. As a result a set of metrics have been developed, which are based on network structure, network dynamics, and network content of communication patterns. Therefore we conceptualize the social aspect of VCoPs on the basis of the three dimensions of social interaction described by Gloor and colleagues (2017b): (i) *connectivity*, where we analyze the network structure; *(ii) interactivity*, where we measure the evolution of the interaction patterns over time; and (iii) *language use*, where we measure the community content, in terms of sentiment, and emotionality.

Moreover we conceptualize the participation as the growth of a community. Although many previous empirical studies used the volume of activity as a proxy of community participation (Malinen, 2015), Iriberri and Leroy (2009) show that the amount of active members involved in the community can be used as a more reliable indicator of community participation and as a proxy for the community success.

## 3. Research Model and Hypotheses

Our research model is presented in Figure 2 and the formulation of the study hypotheses is laid out in the subsequent sections.

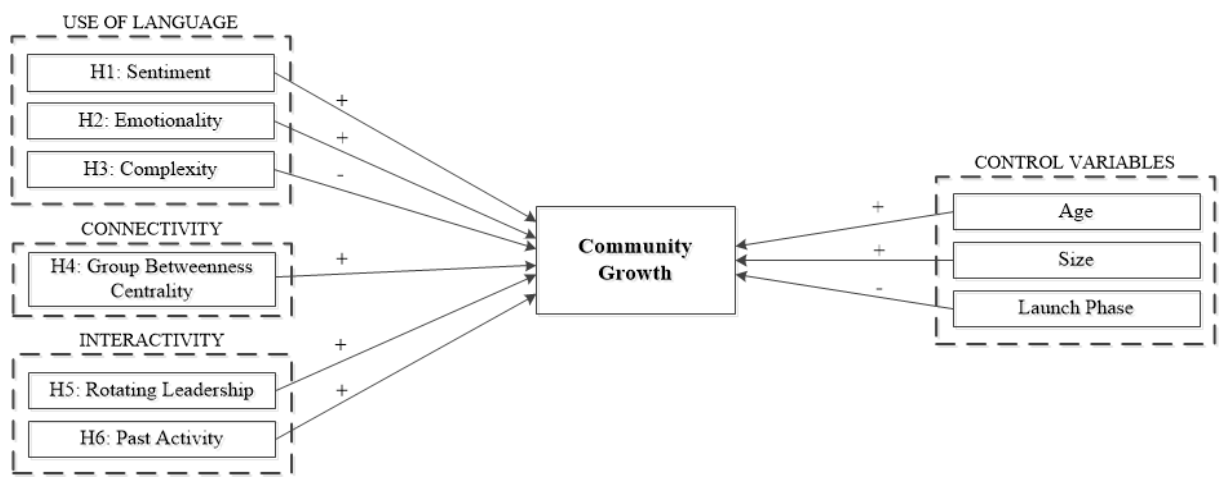

**Figure 2**. Research Model.

*3.1 Use of Language*

Individuals are more willing to share knowledge and engage in virtual communities characterized by high levels of social trust (Huysman and De Wit, 2004; Inkpen and Tsang, 2005; Chiu, Hsu, and Wang, 2006), reciprocity and cooperation (Wasko and Faraj, 2000; Wasko and Faraj, 2005; Chiu, Hsu, and Wang, 2006). Hara and Hew (2007) also demonstrate that members of health-related VCoPs are more likely to voluntarily contribute (self-select) to communities having a non-competitive environment. These findings are grounded on basic principles of human psychology. In communities with a social environment characterised by trust and collaboration, it is likely that members develop positive sentiments toward each other. These positive sentiments increase the willingness of individuals to share knowledge and actively participate in virtual communities, whereas the development of negative affective states increases members' resistance to engage in communities' discussions (Nahl, 2005; Lin, Hung, and Chen, 2009). Accordingly, we hypothesize a positive relationship between the sentiment expressed in the community messages and the participation in the community.

*H1. Communities with members showing a more positive sentiment in the language used will grow faster.*

High levels of personal involvement and emotional attachment increase community members' engagement and lead them to be more willing to exchange opinions, information and ideas (Blanchard and Markus, 2004; Chang and Chuang, 2011). Moreover, members who actively interact and discuss topics in a specific community in cyberspace tend to develop the same emotional attachment to the virtual site as if they were in a physical place (Koh and Kim, 2004). Therefore, we expect that communities where people are more emotionally involved (members deviate from neutral sentiment) will increase the desire for new members to join and actively participate in the community.

*H2. Communities where conversations are more emotional will grow faster.*

The use of a shared language or codes between members increases the social capital of a community (Chang and Chuang, 2011; Tsai and Ghoshal, 1998). Common language is important because it increases the likelihood that members understand each other more easily, thus assisting them in sharing community objectives and behavioural adequacy in VCs (Tsai and Ghoshal, 1998). Sharing a common language fosters the efficiency and efficacy of communication in VCs (Wenger, McDermott, and Snyder, 2002; Chiu, Hsu, and Wang, 2006; Chang and Chuang, 2011). It enables members to gain access to information, helps evaluate the quality of information exchanged, reduces misunderstandings between participants and enhances their capability to catch the knowledge through social exchange (Chiu, Hsu, and Wang, 2006; Chang and Chuang, 2011; Nahapiet and Ghoshal, 1998). Conversely, when members' languages and codes are different, they tend to be isolated and their ability to access information is much more limited (Nahapiet and Ghoshal, 1998). Hence, shared language is considered essential to motivate participants to be more actively involved in the community's life (Chiu, Hsu, and Wang, 2006; Chang and Chuang, 2011). We operationalized the concept of shared language through the measure of Complexity (described in Section 4). Accordingly, we hypothesize that the use of a common language makes VCoPs more attractive.

*H3. Communities in which members use a less complex language (i.e with a smaller vocabulary) will grow faster.*

*3.2 Degree of connectivity*

According to the theories of social capital and collective action, the network structures derived by social interactions between individuals are important predictors of collective and individual actions (Burt, 2009; Putnam, 1995; Rogers and Kincaid, 1981). In particular, the individual position in a network structure of a VCoP, influences the willingness of community members to share knowledge with others (Wasko and Faraj, 2005).

The structural position within a community network represents the social dimension of the identity of CoPs members, which "serves as a pivot between the social and the individual" (Wenger, 1998, p. 145) and can be defined according to their levels of expertise. The identity of CoPs members might range from novice to expert (Lave and Wenger, 1991; Wenger, 1998). The status of "expert" within a community is the result of a negotiation process, which takes place through the interactions and communications in the social environment. Members with a higher expertise are socially recognized as experts (Nistor et al., 2014). Other than to be characterized by superior knowledge and skills, experts within a CoP are also involved in more activities and contribute a higher quality dialogue, thus increasing the level of participation and the attractiveness of the community discussions. Experts, therefore often have a central position and the status of expert in a VCoP can be quantitatively measured trough SNA centrality measures (Nistor et al., 2014, Nistor and Fisher, 2012).

For this reason, we expect that communities with a more centralized communication structure, often suggesting the presence of more major experts, are likely to have a higher growth rate. Network centralization is calculated as the group betweenness centrality score (Freeman, 1977; Wasserman and Faust, 1994).

*H4. Communities with a more centralized communication structure will grow faster.*

*3.3 Degree of Interactivity*

Rotating leadership, revealed by rapidly changing network structures where actors oscillate between peripheral and more central network positions, can have a relevant impact on communication styles. In particular, a democratic leadership style and the adoption of distributed responsibilities, enhance knowledge sharing dynamics in virtual communities and nurtures the engagement level of participants (Kayworth and Leidner, 2002; Lin, Hung, and Chen, 2009). Moreover a rotating leadership style increases the perceived value of joining the community, as it positively affects individual and group creativity and has been found to be a predictor of innovative performance (Gumusluoglu and Ilsev, 2009; Allen et al., 2016). Consistently, we expect that a more democratic participation in the community discourse, with members oscillating between low and high betweenness centrality scores, will increase the attractiveness of the community.

*H5. Communities where actors rotate more – changing their position from being central to being peripheral, and vice versa – will grow faster.*

Communication activity is at the core of the social interaction processes. Activity and responsiveness in the communication process play a vital role in keeping a community alive and a larger knowledge sharing activity leads to a wider active participation (Koh and Kim, 2004). Consistently, Koh and Kim (2004) showed that community knowledge sharing activity predicts both community participation and community promotion. By contrast, a social structure where there is no communication cannot provide benefits for its members (Butler, 2001). Moreover, in public online environments, more active discussions can produce a more intense information exchange between participants, leading to a better indexing on search engines (Van Looy, 2016). This promotes the community and attracts new members.

We therefore expect a positive relationship between the volume of communication and the capacity of a community to retain existing members and attract new joiners.

*H6. Communities with higher levels of activity (numbers of posts per month) will grow faster.*

Finally, community growth can be deeply influenced by its size and age. Larger and older communities can often benefit from the presence of a bigger number of active members who can promote their communities among colleagues, via word-of-mouth. Moreover, since each community is accessible and searchable on the web, being older and with a larger number of posts will improve the indexing of relevant search engines, such as Google (Van Looy, 2016). This may increase the likelihood of attracting new members. New-born communities, on the other hand, will not benefit from past contributions.

Consistently, research findings show that the impact of factors underlying the flourishing of online communities varies according to the distinctive phases characterizing the online community lifecycle (Malhotra, Gosain, and Hars, 1997; Wenger, McDermott, and Snyder, 2002; Andrews, 2002). Therefore, strategies and recommendations to help online communities to succeed should be developed and calibrated on the basis of the specific lifecycle-phase of the online community (Iriberri and Leroy, 2009). For these reasons, we included in our models three control variables, indicating the age and size of a community, as well as the specific stage of its lifecycle.

## 4. Methodology

In order to test the proposed research model and the relative hypothesis, we conducted an empirical research on a set of health professionals' online communities, namely 16 VCoPs, all operating on the same web platform. Communication among members took place in all of the 16 communities, each covering a specific topic of interest. The analysed platform included contents and discussions on the health care delivery and management. It connected over 19.000 members representing 185 countries, thus enabling the exchange and the dissemination of knowledge and information among professionals having different roles in the healthcare system (e.g. policy decision makers, academics, physicians, etc.). The dataset is composed of forum discussion records collected from April 2008 to April 2015, which comprised more than 14,000 community members and about 20,000 community posts. Data collection for each community started from its creation. We analyzed all the data available for the communities operating on the web platform at the time of data collection – accordingly, our study is including the entire population and is not restricted to a random sample.

More specifically, we analyzed the GHDonline's professional virtual communities, created in 2008 with the aim to diffuse knowledge by enabling researchers and practitioners to share all forms of data, expertise, and resources, widely and quickly. Built on this foundation of creating public goods, GHDonline has developed an approach to community management focused on fostering high-quality, dynamic conversations amongst a group of diverse health care professionals. GHDonline membership is free and is open to the public. Members create profiles which can be viewed by other members. This level of transparency enhances accountability among members and helps professional dialogue.

For the analysis, we used Social Network Analysis (SNA) tools and metrics, a well-known set of methods for the analysis of social interactions (Wasserman and Faust, 1994). We represented each community member as a network node, drawing a link between two nodes when one member commented or responded to another one during the online conversation. This allowed us to study the interaction among all members and its evolution over time.

In addition, a content/sentiment analysis was carried out alongside the traditional SNA approaches. We used the social network and semantic analysis software Condor[1] to calculate structural and content metrics and their evolution over time (Gloor 2006; Gloor 2016; Gloor et al., 2017a; Gloor et al., 2017b).

### *4.1 Definition of variables*

We used the number of the new members joining a community as a proxy for the community success, measured in terms of community growth.

*Joiners* is our dependent variable, calculated as the number of new members who join a community each month, contributing at least one post to the online discussion.

As we discussed in Section 3, community growth can be deeply influenced by its size and age, and by the fact of being in a launch phase. Accordingly, we included the following control variables in our models.

*Size* represents the size of a community, measured as the number of members who contributed at least once to the discussion of the community topics.

*Age* refers to the age of the community, expressed in months since its start-up.

*Launch Phase* is a binary control variable meant to identify whether a community is in this phase or in a later life stage. Specifically within our experimental setting, we were able to identify two thresholds that identified communities in the launch phase: when age was 3 months or younger and when members were less than 50.

---
[1] http://www.galaxyadvisors.com/products/

In the remaining part of this section, we present the variables we used as potential predictors of community growth. These variables represent the three dimensions of social interaction in virtual environments, as illustrated in the work of Gloor et al. (2017a). As suggested by these authors, we refer to the research model presented in Section 3, and we include variables to measure: the degree of connectivity, i.e. the social network structure; the degree of interactivity, i.e. the dynamic changes in the network structure over time; and the use of language, i.e. the sentiment, emotionality and complexity of the community posts.

### *4.1.1 Use of language*

Sentiment, emotionality and complexity are important dimensions of the language used, which proved their value in past research. These metrics – calculated by means of the multilingual classifier included in the software Condor – were used, for example, to assess virtual behaviours of employees and optimize communication styles, with the objective of improving customer satisfaction (Gloor et al., 2017a); or to forecast turnover intentions and disengagement of company managers (Gloor et al., 2017b). As discussed in Section 3, we hypothesize that a less complex and more positive and emotional language will support community growth.

*Sentiment* represents the sentiment of the language used in the community posts. It varies in the range [0,1], where 1 is the most positive value and values below 0.5 indicate a negative sentiment. This variable was calculated using a multi-lingual classifier based on a machine learning method trained on large datasets extracted from Twitter (Brönnimann, 2014).

*Emotionality* is a variable that measures the level of emotionality of the language used, calculated as the standard deviation of sentiment (Brönnimann, 2014). In a community where the level of emotionality is high, there is a more vivid debate in which the positive and negative sentiment frequently alternate (Grippa et al., 2014).

*Complexity* represents the average complexity of the vocabulary used and is calculated as the likelihood distribution of words within a forum post – i.e., the probability that each word of a dictionary appears in the text (Brönnimann, 2014). When the language is more shared, its complexity is lower, indicating that the same set of words is more frequently used among the community members, even if these words could be classified as complex or rare.

### *4.1.2 Degree of Connectivity*

Degree of connectivity reflects the social structure of a community network and is therefore analysed using metrics from Social Network Analysis. To evaluate the overall network structure, we chose Group Degree Centrality, among the many possible centrality measures (Freeman, 1979). Our choice depends on the fact that this metric provides us with an important indication about how much dominated a social network is. It describes whether the network is more or less dependent on overly connected nodes, which more frequently lie in the paths that interconnect the other social actors. Group Betweenness Centrality is therefore a useful proxy for the identification of community leaders and, in past research, it has been correlated to community success and creativity of community members (Kidane and Gloor, 2007).

*Group Betweenness Centrality* indicates how centralized is the communication within a community. The higher the group betweenness centrality, the more the information revolves around few key people – or hubs, or moderators– that have a major role in the network and contribute to keeping different subgroups together. A perfectly dominated network would look like a star, with one central social actor and all the others just interacting with her/him (Wasserman and Faust, 1994). In general, we can consider the community networks as oriented graphs of n nodes (user accounts) – referred as G = {$g_1$, $g_2$, $g_3$ … $g_n$} – and of m directed arcs (community posts) linking these nodes, when one user responds to another. Individual betwennes centrality scores are calculated as suggested by Wasserman and Faust (1994):

$$B_C(g_i) = \sum_{j<k} \frac{d_{jk}(g_i)}{d_{jk}}$$

Where $d_{jk}$ is the number of shortest paths linking the generic couple of nodes $g_j$ and $g_k$, and $d_{jk}(g_i)$ is the number of that paths which contain the node $g_i$. This measure can be standardized dividing it by [(n –1) (n – 2) / 2]. The group betweenness centrality score, is then given by:

$$GB_C = \frac{2\sum_{i=1}^{n}[B_C(g^*) - B_C(g_i)]}{[(n-1)^2(n-2)]}$$

Where $B_C(g^*)$ is the largest $B_C(g_i)$ score for the actors in the group.

### *4.1.3 Degree of interactivity*

In this study, degree of interactivity is represented, for a first part, from past activity of community members, which can enrich the knowledge shared in a VCoPs and be a determinant of attraction for new members; for a second part, by the oscillations in betweenness centrality for the members of a community (rotating leadership). A communication style based on rotating leadership proved to favour knowledge sharing and to foster innovative performances (Allen et al., 2016). Having high scores of rotating leadership means that a community is operating more "democratically", being less dominated by few static leaders. In such a community one could expect a more free and dynamic knowledge sharing activity, which can attract more community members (Davis and Eisenhardt, 2011; Kidane and Gloor, 2007).

*Past Activity* is the number of messages posted in a community in the month that precedes the current one.

*Rotating Leadership* is measured as the average number of oscillations in betweenness centrality for the members of a community each month (Davis and Eisenhardt, 2011). Specifically, considering the individual betweenness centrality scores of the members of a group, rotating leadership counts the number of times individual scores reached local maxima or minima. It can also be interpreted as a measure of the liveliness of a discussion, reflected in network structures that change frequently, with actors fluctuating between peripheral and central network positions (Kidane and Gloor, 2007).

## 5. Results

In our models, we study – longitudinally – the influence of content and network metrics on the number of new active members. The communities we analysed are homogeneous with respect to the target audience and to the web platform used, so we do not need to control for such differences. In addition, none of the analysed communities received external promotion through a specific marketing campaign. Attraction of new members was mainly based on word-of-mouth, with the additional possibility to find the communities by crawling the web: contents are publicly available and indexed by the major search engines.

In the empirical data analysis, we first examined the correlation among the abovementioned variables, then we tested the influence of predictor variables on the number of joiners through a multi-level regression model, in order to take into account the nested nature of data (Fidell, Tabachnick, and Linda, 2007; Young, 2013). Pearson's correlation coefficients are presented in Table 1.

|   |   | 1 | 2 | 3 | 4 | 5 | 6 | 7 | 8 | 9 | 10 |
|---|---|---|---|---|---|---|---|---|---|---|---|
| 1 | Joiners | 1 | | | | | | | | | |
| 2 | Age | .195** | 1 | | | | | | | | |
| 3 | Size | .292** | .889** | 1 | | | | | | | |
| 4 | Launch Phase | -.170** | -.575** | -.530** | 1 | | | | | | |
| 5 | Emotionality | -.080* | .047 | .013 | -.065 | 1 | | | | | |
| 6 | Sentiment | .063 | .055 | .105** | -.053 | -.012 | 1 | | | | |
| 7 | Complexity | -.229** | -.244** | -.265** | .129** | .046 | -.202** | 1 | | | |
| 8 | Past Activity | .300** | .261** | .396** | -.233** | -.082* | .053 | -.167** | 1 | | |
| 9 | Group Betweenness Centrality | .374** | .147** | .149** | -.045 | -.056 | -.131** | -.325** | .171** | 1 | |
| 10 | Rotating Leadership | .234** | .086* | .132** | -.104** | -.056 | .002 | -.122** | .219** | .146** | 1 |

**p<0.01; *p<0.05.

**Table 1.** Community growth: correlation coefficients (N=754).

Table 1 shows that the number of new members is correlated with a lower complexity of the language (initially supporting H3), but not significantly with sentiment and emotionality. Therefore, it seems that the use of a common language – regardless of its sentiment and emotionality – is beneficial to community building. We posit this result is also influenced by the topics covered by each community, here all meant for a specific audience of healthcare professionals. With regards to structural and interaction metrics, we found that both group betweenness centrality and rotating leadership are positively correlated with the number of new joiners, thus supporting H4 and H5 and pointing to the importance of the role of informal leaders within the community. New members will be attracted by a community where some key people act as connectors with different subgroups and contribute significantly to the online discussion, sharing their knowledge and advice. In addition, it seems that newcomers prefer communities where rotating leadership is higher, with more members fluctuating frequently from central to peripheral discussions, not holding a static position. We also found a significant correlation between joiners and past activity levels, as expected in H6. This is probably due to the greater visibility that a community acquires when discussions become more active. Moreover there is a positive association between size and past activity, probably due to the fact that more messages are usually exchanged when there are more members in a community and when the community gets older. On the other hand, size and activity are significantly smaller in the launch phase. Additionally, age significantly correlates with size and joiners, once again confirming the greater attractiveness of older communities. Our main control variables – age, size and launch phase – are correlated to one another thus leading to potential collinearity problems. Also the conceptual link among these three variables is strong: we observed that communities that are older, bigger and not anymore in the launch phase attract more new members – supporting the concept of preferential attachment, where older and more connected nodes show a higher attractive power (Jeong, Neda, and Barabasi, 2003). One solution to collinearity was performing a principal component factoring, combining our control variables and retaining just one single factor, accounting for 80% of the variance, with the following factor loadings: .94 for age, .93 for size and -.77 for launch phase. This resulted in an additional factor variable which we called *Maturity,* being higher for communities that are older and of a bigger size and smaller when communities are in the launch phase. We used this new control variable in the multilevel regression models presented in Table 2. When dealing with repeated measures over time, using a mixed model is more appropriate than using multiple regression models: it allows random intercepts and slopes, to control for significant differences between communities.

| Variable | Model 1 | Model 2 | Model 3 | Model 4 | Model 5 |
|---|---|---|---|---|---|
| **Constant** | 6.054** | 6.252** | 28.071** | 1.697** | 7.385** |
| **Maturity** | | 1.137** | | | |
| **Sentiment** | | | -2.126 | | |
| **Complexity** | | | -2.892** | | -.846* |
| **Emotionality** | | | -7.294 | | |
| **Past Activity** | | | | 0.039** | 0.038** |
| **Group Betweenness Centrality** | | | | 6.029** | 5.730** |
| **Rotating Leadership** | | | | 1.951** | 1.914** |
| **Variance Level 2** | 2.871 | 2.252 | 2.381 | 1.836 | 1.748 |
| **Variance Level 1** | 24.196 | 23.184 | 23.292 | 19.458 | 19.340 |
| **ICC** | 10.61% | | | | |
| **Change in variance Lev. 2** | | -21.56% | -17.07% | -36.02% | -39.11% |
| **Change in variance Lev. 1** | | -4.18% | -3.74% | -19.58% | -19.82% |
| **N** | 754 | 754 | 754 | 754 | 754 |
| **Groups** | 16 | 16 | 16 | 16 | 16 |

**p<0.01; *p<0.1.

**Table 2.** Determinants of community growth.

Results from the multilevel regression models (Table 2) support hypotheses H4, H5 and H6, with a relevant effect of group betweenness centrality over the other predictors. These findings give evidence to the importance of having formal or informal leaders in a community who can support and facilitate the discussion; the community attractiveness will be even higher when leaders are not static, but rotate more supporting a more distributed participation. Past activity is also a relevant driver of community growth, supporting the idea that a more active community is more attractive and potentially more popular on the web. H3 is supported by our models, even if the effect of language complexity is much smaller when considered together with the other predictors. Sentiment and emotionality, on the other hand, do not contribute significantly in this context. Lastly, our findings are reinforced by an intraclass correlation coefficient of 10.61% which suggests that only a small part of variance is accountable to the differences between the communities.

Adding random slopes to the models did not significantly improve them. We also controlled for seasonal trends with no big relevance for our models. We only noticed a significant negative effect of Christmas holidays on the number of new members. Figure 3 summarizes our results.

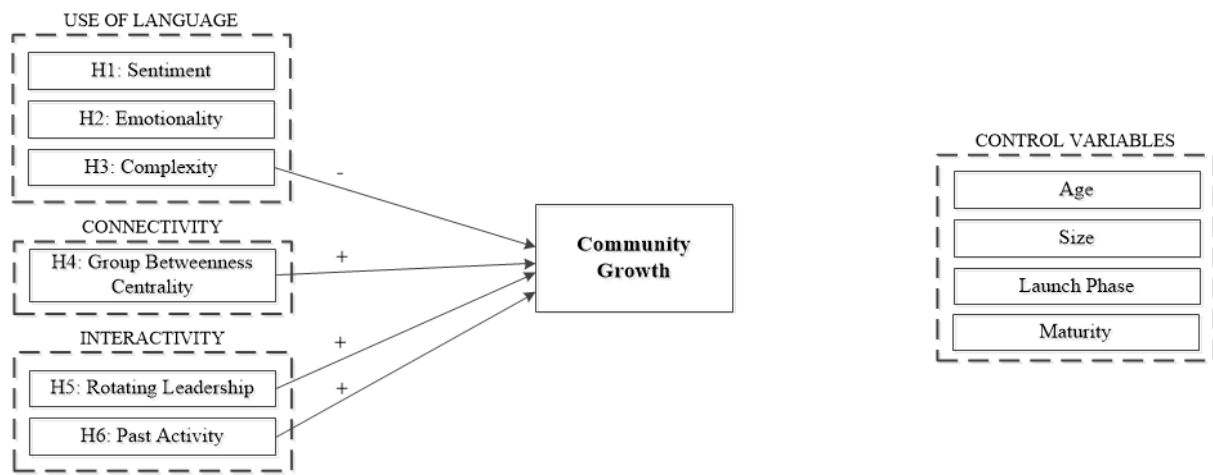

**Figure 3**. Significant Predictors of Community Growth.

# 6. Discussion and Practical Implications

According to the results presented in Section 5, the most important factors fostering the growth of VCoPs for healthcare professionals are: the level of past activity, the use of a less complex language, the presence of a centralized leadership structure and of rotating leaders frequently changing their position over time.

As expected, our findings show that communities where people interact more, attract more members. This result confirms previous studies exploring the dynamics of virtual communities (Koh and Kim 2004; Butler, 2001) and suggests a virtuous circle, where an increase in the knowledge sharing activity leads to a greater participation and so forth.

Consistently with the theories of collective action and social capital (Burt 2009; Putnam 1995), our study also reveals that the growth of a VCoP is influenced by the behaviours of people playing a major role in the community. Activities in CoPs and VCoPs are informal, highly voluntary, and they heavily rely on members' self-motivation and initiative. Informal leadership structure, other than formal structure, plays therefore a pivotal role in the success of the community. Inside CoPs, informal communities' leadership is indeed considered one of the most critical success factors to the increase of activity and participation to the community (Wenger, 1998; Zboralski, 2009). Individuals who are centrally embedded in a social network of a community are more likely to cooperate and share knowledge with others (Wasko and Faraj, 2005). Moreover, community members assume a central position within a network as result of a negotiation process within the social environment, in which they are recognized as experts by other participants (Nistor et al., 2014). People who are embedded in the center of a social network of a VCoPs are likely to have a greater expertise in the field discussed compared to other members and therefore they are more likely to provide a greater contribution in terms of quality of the discussion. This explains our result about the positive link between network centralization and attractiveness of a community. These findings are also aligned with previous research highlighting the importance of moderators and people with high centrality scores, on the overall members' participation (e.g. Wasko and Faraj, 2005; Hara and Hew, 2007; Barnett et al., 2012; Nistor et al., 2014; Baek and Kim, 2015).

We also found that VCoPs having a network structure which rapidly changes over time, with rotating leaders, are likely to attract more active members. This result is aligned with previous research showing how a "democratic" leadership style positively contributes to the improvement of knowledge sharing dynamics within VCoPs, thus positively affecting members participation (Kayworth and Leidner, 2002; Lin, Hung, and Chen, 2009).

The fact that a VCoP grows more when its structure is more reliant on the communication of central people and when there is more rotating leadership, can be explained by the importance played by people having the role of "experts" within high-specialized communities. In fact, in this type of communities the presence of "experts" on the specific topic being discussed is extremely beneficial to the value of the debate (Wasko and Faraj, 2005; Nistor et al., 2014).

In addition to factors related to the network structure and dynamics, our findings provide some evidence that the use of a less complex, more shared, language in the communication between communities' members contributes to the growth of VCoPs. These results are aligned with previous studies pointing to the importance of a shared language as a key enabler of knowledge-sharing dynamics within VCoPs (Wenger, McDermott, and Snyder, 2002; Chiu, Hsu, and Wang, 2006; Chang and Chuang, 2011). This holds particularly true in our study focusing on VCoPs in the healthcare setting. In fact, like for other highly specialized communities, discussions between healthcare professionals are characterized by specific conventions of language, which are known only by the experts of the field in question (Diesner and Evans, 2015). However, our findings also indicate that, although the use of a shared language is important for the growth of a VCoP, it is not as important as the network structure and dynamics.

Chiu, Hsu, and Wang, (2006) suggest that shared language can increase the quality of shared knowledge, but not its quantity. However, when a language is too limited – with the members of a group repeatedly discussing the same topic, without adding new knowledge, this could hinder the creation of new concepts and the introduction of new ideas.

Finally, our results show that the expression of positive (or negative) sentiment and intense emotional states in the messages exchanged by community members, do not have a significant impact on the growth of VCoPs. This results contrasts with previous studies maintaining the beneficial impact of a positive sentiment (Nahl, 2005; Lin, Hung, and Chen, 2009) and emotionality (Blanchard and Markus, 2004; Chang and Chuang, 2011) on community success. However, professional communities are indeed characterized by slightly positive sentiments and low emotionality, given the professional conventions of the language used (Diesner and Evans, 2015). Compared to other types of VCs, within VCoPs for healthcare professionals it is unlikely that practitioners express strong negative or positive feelings towards colleagues, especially in public discussions (Chmiel et al., 2011). This is also evident in our experiment where average sentiment is positive, but slightly above the neutral and with a low standard deviation (M 0.57, SD 0.11); similarly, we found a low average emotionality with a very low standard deviation (M 0.27; SD 0.02), proving that the sentiment of the messages is almost always positive or neutral.

This study contributes both to the advancement of literature and practice in the field of online community building. It represents probably the first attempt to analyse the impact of social aspects on the growth of VCoPs taking simultaneously into account the joint effect of the three dimensions of social interactions (use of language, connectivity, interactivity). In addition we investigated the effects of new promising metrics such as the betweenness centrality oscillations (rotating leadership). While the level of past activity is commonly recognized as an important contributor to the growth of VCoPs, the role of rotating leadership in supporting the interaction and creating a more fluid communication environment is a new factor, which has not previously investigated by other studies.

Compared to other studies using qualitative methods (e.g. surveys, interviews) – which prevail in the field and are mostly descriptive (Bolisani and Scarso, 2014) – we offer an overview of an analytical approach that could be easily replicated by community managers and practitioners interested in how to improve the management of VCoPs.

As the role of healthcare VCoPs as an important means for knowledge sharing and professional support between healthcare professionals is increasingly recognized (Rolls et al., 2008; Parboosingh, 2002; Winkelman and Choo, 2003, Ho et al., 2010), the identification of success drivers and practical strategies potentially leading to their flourishing should not be overlooked. Dholakia, Bagozzi, and Pearo (2004, p. 261) suggest that virtual communities are "only likely to grow in importance, influence, and the activities for which they are used as consumers become more comfortable and acclimatised to these environments and marketers learn how to forecast, monitor and design their communication programs to take advantage of such opportunities". The way an online community is managed, decisively influences the willingness of members to contribute to the community (Iskoujina and Roberts, 2015).

Practical implications for community managers are twofold. First, we confirm previous research pointing to the importance of social interactions to the success of VCoPs (Amichai-Hamburger et al., 2016; Jeon, Kim, and Koh, 2011; Hara and Hew, 2007; Nistor et al., 2014).

Second, the perspective of analysis adopted in this study as well as the choice of relative metrics, have been purposefully designed to provide managers of VCoPs with relevant information, which can be easily derived from data available on the online platforms. Compared to qualitative approaches of studying social interactions within VCs (Malinen, 2015), the quantitative approach proposed in this research supports the development of more data-driven analytic tools.

On the basis of these insights platform managers could, for example: monitor past activity and identify attractive discussion topics, split or merge existing communities on the basis of their similarities in the language used or in the network structure; most importantly, they could better measure the role of experts and elect formal moderators, or recruit participants with high skills in a specific knowledge domain.

Lastly, our findings are particularly relevant for practitioners interested in starting new online communities involving healthcare professionals; our insights motivate further research testing our methods and tools in other knowledge intensive industries or professional contexts.

## 7. Limitations and Conclusion

This study has identified factors leading to the growth of healthcare VCoPs using a new quantitative approach to the analysis of social interactions in this setting, taking simultaneously into account the content, the structure, and the longitudinal dynamics of community communication.

One of the major findings of this study is that, differently from what is reported for other types of online communities, within VCoPs emotionality and sentiment of the messages exchanged are less important than structural and dynamic aspects of the network - the growth of VCoPs is mostly related to the presence of leaders/experts and to their ability to rotate to foster a more democratic participation. .

The fact that some of the variables identified in this study significantly impact the growth of the communities, also suggests that the use of analytics tools to monitor social dynamics within VCoPs is a promising approach for increasing their success.

Limitations of this study and directions for future research are presented here. First, we focused on 16 VCoPs for healthcare professionals hosted by the same platform; specifically we analysed all the data available on the platform at the time of data collection. The 16 analysed communities are representative of a range of different healthcare topics sometimes pretty different, such as HIV prevention or Global Surgery. Moreover the platform is targeted to a variety of audiences, with participants sometimes carrying out very different jobs in the healthcare context (e.g. nurses, physicians, accountants), across different countries. Even if the problem of differences between communities is partially solved by the use of multilevel regression models (which also show a rather low intraclass correlation coefficient), we encourage future research on a larger number of online communities operating in knowledge-intensive sectors other than healthcare. Moreover, the inclusion of more control variables – such as individual traits (e.g. age, location, personality and web literacy) or professional characteristics of members (e.g. domain knowledge, experience, job, organization and formal role) - could help verify and generalize our findings.

Lastly, although compared to other on VCoPs adopting mainly SNA variables (Zaphiris and Sarwar, 2006; Pfeil and Zaphiris, 2009; Stewart and Abidi, 2012; Nistor et al., 2014) our analysis has been enriched by content and interaction metrics, we did not consider more qualitative aspects such as the evolution of specific discussion topics. Therefore, in line with Malinen (2015) we suggest further research to combine both qualitative and a quantitative approaches.

Chang, H. H., and Chuang, S. S. (2011), "Social capital and individual motivations on knowledge sharing: Participant involvement as a moderator", *Information and management*, Vol. 48 No. 1, pp. 9-18.

Chen, I. Y. (2007), "The factors influencing members' continuance intentions in professional virtual communities-a longitudinal study", *Journal of Information Science*, Vol. 33 No. 4, pp. 451-467.

Cheung, C. M., Lee, M. K., and Lee, Z. W. (2013), "Understanding the continuance intention of knowledge sharing in online communities of practice through the post-knowledge-sharing evaluation processes", *Journal of the American Society for Information Science and Technology*, Vol. 64 No. 7, pp. 1357-1374.

Chiu, C. M., Hsu, M. H., and Wang, E. T. (2006), "Understanding knowledge sharing in virtual communities: An integration of social capital and social cognitive theories", *Decision support systems*, Vol. 42 No. 3, pp. 1872-1888.

Chmiel, A., Sobkowicz, P., Sienkiewicz, J., Paltoglou, G., Buckley, K., Thelwall, M., and Hołyst, J. A. (2011), "Negative emotions boost user activity at BBC forum" *Physica A: statistical mechanics and its applications*, Vol. 390 No. 16, pp. 2936-2944.

Constant, D., Sproull, L., and Kiesler, S., (1996) "The kindness of strangers: The usefulness of electronic weak ties for technical advice", *Organization science*, Vol. 7 No. 2, pp. 119-135.

Cox, A. (2005), "What are communities of practice? A comparative review of four seminal works", *Journal of information science*, Vol. 31 No. 6, pp. 527-540.35